\documentclass{acm_proc_article-sp}
\usepackage{graphicx}
\usepackage{mathtext}
\usepackage{amssymb}

\begin{document}

\title{What type of distribution for packet delay in a global network should be used in the control theory?}

\numberofauthors{2} %  in this sample file, there are a *total*
\author{
% 1st. author
\alignauthor
Andrei M. Sukhov\titlenote{corresponding author}\\
       \affaddr{Samara State Aerospace University}\\
       \affaddr{Moskovskoe sh., 34}\\
       \affaddr{Samara, 443086, Russia}\\
       \email{amskh@yandex.ru}
\and
% 2nd. author
\alignauthor 
Natalia Kuznetsova\\
       \affaddr{Samara State Aerospace University,\\ 
       Togliatti branch}\\
       \affaddr{Voskresenskaya st., 1}\\
       \affaddr{Togliatti, 445000, Russia}\\
       \email{meneger\_job@mail.ru}
}

\date{July 26, 2009}

\maketitle

\begin{abstract}
In this paper correspondence between experimental data for packet delay and two theoretical types of distribution is investigated. Calculations have shown that the exponential distribution describes the data on network delay better, than truncated normal distribution. Precision experimental data to within microseconds are gathered by means of the RIPE Test Box. In addition to exact measurements the data gathered by means of the utility {\em ping\/} has been parsed that has not changed the main result. As a result, the equation for an exponential distribution, in the best way describing process of packet delay in a TCP/IP based network is written. The search algorithm for key parameters as for normal, and an exponential distribution is resulted.
\end{abstract}


 % NOT required for Proceedings

\section{Introduction}
\label{intr}

The special area of the control theory, named networked control systems in which transfers as environment of operating signals were used computer networks, has arisen in the late nineties of the XX-th century~\cite{zbp}. Originally, as the network environment of control systems local networks~\cite{Georges05} which differ high-speed data transfer and in the minimum percent of packet loss were used.

Framing of criteria of stability for networked control systems in which as the handle environment the global network Internet is used, is extremely complicated because of random character of distribution of packet delay and their big absolute values~\cite{Hespanha07,tch,Zampieri08}. Non-use of knowledge of network processes, including measurement methods of an available bandwidth, knowledge of types of distribution of packet delay, methods on compensation of packet losses is an obvious obstacle in path of development of networked control systems.

However till now, results of the advanced network researches are not used in the control theory, no less than algorithms on their basis are not created. The present project assumes introduction of new network decisions in networked control systems.

For the decision of problems of the networked control systems on the basis of stack TCP/IP it is more convenient to use the process approach to the control theory, based on idea of existence of some universal functions of control. The purpose of our research is the finding of these function for network components. In the modern theory of computer networks there were many utilities working with a network delay, there is a progress in studying and modelling of transmiting of packages. Our problem consists in trying to describe process of a network delay of management packages and to show ways of practical calculation of all parametres entering into corresponding distribution functions~\cite{Fridman04}.

By transmission of control signals through TCP/IP etwork, the separate packages of the controlling data flow transferring the information, are supplied non-uniformly, and the part of packages in general is lost by transmission on a network and does not reach a target. For rise of efficiency of control algorithms it is necessary to reduce to a minimum of packets delay and their variation, and also percent of packet loss. Similar algorithms are used for transmission voice and video streams, in grid systems, at control of robust systems, in network computer games, etc.

At first it would be desirable to result the brightest research on a distribution type for network delay. To understand, about what there is a speech in described papers, will give definitions of notations used in them:

\begin{itemize}
	\item 
	Round-trip time ({\em RTT\/}) time is the time required for a packet to travel from the testing host to a the remote computer that receives the packet and retransmits it back to the source.
	\item
	The One-Way Delay ({\em OWD\/}) value is calculated between two synchronized points A and B of an IP network, and it is the time in seconds that a packet spends in travelling across the IP network from A to B.
\end{itemize}

In particular, Elteto and Molnar~\cite{Elteto99} have spent measurements of  round-trip delay  in in the Ericsson Corporate Network, complex analysis of the received data has allowed to build the supposition about distribution type for network delay. The main finding of their research is that the round-trip delay can be well approximated by a truncated normal distribution.

Konstantina Papagiannaki {\em et al\/}~\cite{pmft} in the research have measured and have analyzed packet delay between two adjacent routers in the core network. On the basis of the received measurements, they have made the supposition about the factors influencing occurrence of delay, and very big delays which cannot be explained in the way of batch processing in routers on algorithm FIFO have been noticed.

Recently, fulfilling a series of operations on measurement of an available bandwidth~\cite{sss}, we have installed that for a type definition of delay distribution we should research only a variable part of delay while its most part remains constant. This fact also has served as a starting point of our operation.

\section{Premises for model}
\label{sb}

In 1999 Downey \cite{dow} for the first time has detected linear dependence of the minimum possible round trip time on the size of transferred packets.
In 2004 precise experiments by Choi et al \cite{chm} proved that the minimum fixed delay component $D^{fixed}(W)$ for a packet of size $W$ is a {\it linear} (or precisely, an {\it affine}) function of its size, 
\begin{equation}
  D^{fixed}(W)=W\sum_{i=1}^h 1/C_i + \sum_{i=1}^h \delta_i
  \label{eq2}
\end{equation}
where $C_i$ is each link of capacity of $h$ hops and $\delta_i$ is propagation delay. To validate this assumption, they check the minimum delay of packets of the same size for three path, and plot the minimum delay against the packet size. 

Let $D(W)$ represents the point-to-point delay of a packet. Here we refer to it as the minimum path transit time for the given packet size $W$, denoted by $D^{fixed}(W)=\min D(W)$. With the fixed delay component $D^{fixed}(W)$ identified, we can now substract it from the point-to-point delay of each packet to study the variable delay component $d^{var}$. The variable delay component of the packet, $d^{var}$, is given by 
\begin{equation}
	D(W)=D^{fixed}(W)+ d^{var}
	\label{dvar}
\end{equation}
 
Computed minimal delay $D^{fixed}(W)$ is
\begin{equation}
	D^{fixed}(W)=D_{min}+W/C,
	\label{C-for}
\end{equation}
where $C$ is end-to-end capacity and
\begin{equation}
	D_{min}=\lim_{W\rightarrow 0}D^{fixed}(W)
	\label{Dmin}
\end{equation}
The value $D_{min}$ is related to the distance between the sites (i.e. propagation delay) and per-packet router processing time at each hop along the path between the sites \cite{cml,cc}. This value represents as the minimum delay $D_{min}$ for which the very small package can be transmitted on a network from one point in another. 

The minimal delay~\cite{sss} of datagram transmission $D_{min}$ may be calculated as
\begin{equation}
  D_{min}=\frac{W_2D_1 - W_1D_2}{W_2-W_1}
\end{equation}	
This value as well as the methods of its measurement has a important significance in applied tasks of control theory~\cite{zbp}. The second significant question of networking control theory is the distribution type for variable delay component $d^{var}$ which should be studied. To know the expression for this parameter we may easy calculate the duration of buffer for streaming aplication on receiving side.

\section{Experimental search}
\label{s2}

To determine distribution type for a variable delay component $d^{var}$ we should gather enough considerable quantity of measurements between various hosts in the Internet, made with a precise accuracy. The basic problem of experimental testing is the precise of delay measurements that is necessary for accurate result. The exact metering demands micro second precision for delay measurements; we are reaching such accuracy with help of RIPE Test Box mechanism~\cite{ggk,ripe}. In order to prepare the experiments three Test Boxes have been installed in Moscow, Samara and Rostov on Don during 2006-2008 years in framework of RFBR grant 06-07-89074. Each RIPE Test Box represents a server under management of an FreeBSD operating system with the GPS receiver connected to it.

Characteristic times of investigated processes (a packet delay, jitter) have the order from 10 $ms$ to 1 $sec$, therefore is quite enough accuracy of system hours of a RIPE Test Box for their reliable measurement. At the first stage experiment between tt01.ripe.net (RIPE NCC at AMS-IX, Amsterdam), tt143.ripe.net (Samara, SSAU), tt17.ripe.net (Bolonia) and tt74.ripe.net (Melburn) have been made which included precision measurement of packet delay with accuracy 2-12 $\mu s$. Testing results are available in telnet to corresponding RIPE Test Box on port 9142. It is important to come and write down simultaneously the data on both ends of the investigated channel. 

On the basis of the received data set it is easy to construct a cumulative distribution function for network delay $D$:
\begin{equation}
	F(D)=P(x\leq D)
	\label{del}
\end{equation}
 
For initial comparison truncated normal and exponential distributions have been chosen, expressions for which are written down.

For truncated normal distribution it is possible to select following approximation:
\begin{equation}
F(D)=\left\{\begin{aligned}
0, \quad D\leq D_{min} ;\\
\frac{\sqrt{2/\pi}}{\sigma}\int\limits_{D_{min}}^D \exp \left\{-\frac{(x-D_{min})^2}{2\sigma^2}\right\}dx,\\ \quad D>D_{min}
\end{aligned} \right.
\label{Dnor}
\end{equation}
where 
\begin{equation}
\sigma=D_{av}-D_{min}
\label{sigma}
\end{equation}
 is the difference between average network delay $D_{av}(W)=\mathbb{E}[D(W)]$ and  minimum delay $D_{min}(W)$.

It is necessary to mark that all statistics was gathered by us for the fixed size of a packets $W$. By default for RIPE Test Box it equals to 100 bytes. Later we update a cumulative distribution function $F(D,W)$ taking into account the packets size $W$.

The alternative type of allocation which will be checked on correspondence is an exponential distribution, expression for which is written below. 
\begin{equation}
F(D)=\left\{\begin{aligned}
0, \quad D\leq D_{min} \\
1-\exp\left\{-\lambda (D- D_{min})\right\},\quad D>D_{min}
\end{aligned} \right.
\label{Dexp}
\end{equation}
where 
\begin{equation}
\lambda=1/(D_{av}-D_{min}) 
\label{lambda}
\end{equation}
is reciprocal to the difference between average network delay $D_{av}(W)=\mathbb{E}[D(W)]$ and  minimum delay $D_{min}(W)$.

For check of conformity to distribution type two methods will be used: calculation of Pearson correlation coefficient and a graphic method. We will designate as $K_{nor}$ correlation coefficient between experimental and normal distributions then $K_{exp}$ is correlation coefficient between experimental and exponential distributions. 

The data obtained by us is shown in Table~\ref{prD}, where the column {\bf host} corresponds to a direction between two RIPE Test Boxes, and the column $W$ specifies in the size of a testing packet.

\begin{table}
	\centering
		\begin{tabular}{|l|l|c|c|c|} \hline
N & host & $W$ (bytes) & $K_{nor}$ & $K_{exp}$ \\ \hline
1 & bolonia &&&\\
  & tt01->tt17 & 100 &0.76 & 0.97 \\ \hline
2 & samara    &&& \\
  & tt01->tt143 & 100 & 0.87 & 0.98 \\ \hline
3 & samara   &&& \\
  & tt01->tt143 & 1024 & 0.99 & 0.99 \\ \hline
4 & melburn  & && \\
  & tt01->tt74 & 100 & 0.66 & 0.97 \\ \hline  
	\end{tabular}
	\caption{Precise measurements}
	\label{prD}
\end{table}

\begin{figure*}
\centering
\includegraphics[height=5cm]{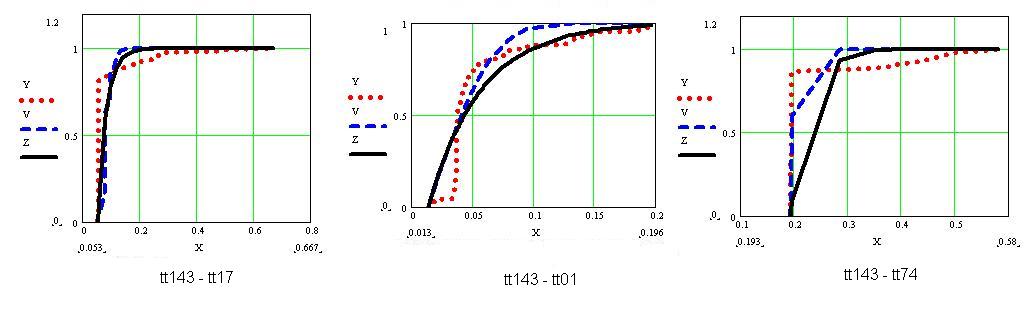}
\caption{Experimental (red), normal (blue) and exponential (black) cumulative distribution function, precise testing}
\label{f1}
\end{figure*}

Except correlation coefficients it is possible to compare and graphics representation of cumulative distribution functions, representing all three functions on one schedule. On the uniform graphics (see Figure~\ref{f1}) red color selects an experimental curve, blue color corresponds to normal allocation. In black colour the exponential distribution is painted.

All experiments resulted above testify that the best type of distribution describing packet delay in a global network, represents an exponential distribution. Thus, as have shown our researches, the random variable of packet delay between two network points is arranged on an exponential low with the parameter calculated from experimental values under the Equation~(\ref{lambda}).

\section{The elementary check}
\label{s4}

However, it is not each investigator who is engaged in the control theory, has at the instruction the precision measuring system, similarly RIPE Test Boxes. Therefore in this part it would be desirable to test delay distributions, leaning against the data of well-known utilities which are not demanding the expensive equipment.

For testing it is possible to use the utility {\em ping} as it is the most widespread and readily available resource for connection quality check in TCP/IP networks. Let's mark only that this utility measures round-trip time, instead of one way delay, correspondence between these values approximately equally $OWD\approx RTT/2$.

We wish to be convinced that the data received {\em ping}, is exact enough that it was possible to judge delay distribution. Using the utility{\em ping}, we have tested connection between points AIST - New Zealand (tt47.ripe.net), Volgatelekom - Australia (tt74.ripe.net) and SSAU-Melbourn (tt74.ripe.net). As remote hosts were used servers of RIPE measurement system, AIST, Volgatelecom and SSAU is local internet Service Providers from Samara region, Russia. Processing the obtained data on the above described algorithm, we have received the results presented in the Table~\ref{pingD}.

\begin{table}
	\centering
		\begin{tabular}{|l|l|c|c|c|} \hline
N & host & $W$ (bytes) & $K_{nor}$ & $K_{exp}$ \\ \hline
1 & AIST &&&\\
  & New Zeland & 32 & 0.94 & 0.95 \\ \hline
2 & Volgatelecom    &&& \\
  & Australia & 32 & 0.96 & 0.98 \\ \hline
3 & SSAU  &&& \\
  & Melburn & 64 & 0.66 & 0.97 \\ \hline
4 & Infolada  & && \\
  & Athens & 32 & 0.98 & 0.98 \\ \hline  
	\end{tabular}
	\caption{{\em ping} measurements}
	\label{pingD}
\end{table}

\begin{figure*}
\centering
\includegraphics[height=5cm]{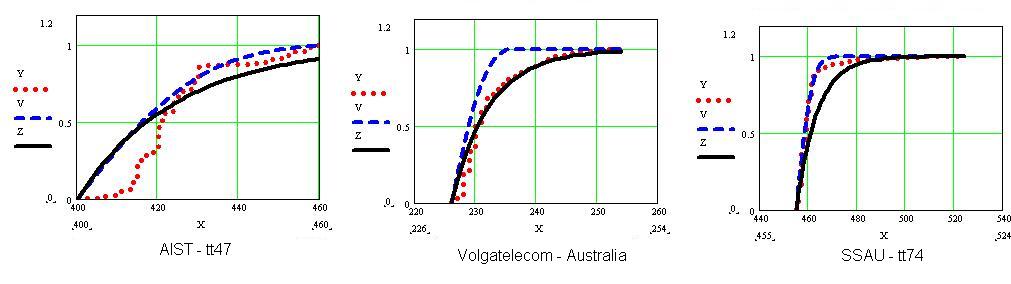}
\caption{Experimental (red), normal (blue) and exponential (black) cumulative distribution function, {\em ping\/} testing }
\label{f2}
\end{figure*}

The evident illustration is resulted in definition of distribution type in Figure~\ref{f2}.

It should be noted that the utility {\em ping\/} allows finding automatically values of variables $D_{av}$ and $D_{min}$ (see Eqns.~(\ref{Dnor}) and (\ref{Dexp})) which completely define the distribution form, both normal and exponential types. It is enough to give sequence from 10 packets to obtain the given values with a split-hair accuracy, sufficient for the description of processes of the control theory.

\section{Distribution type for delay}
\label{s5}

In real the Internet processes the size of transferred packages can vary, therefore the cumulative distribution function should be updated. For each size of a packet $W$ there is the minimum time $D^{fixed}(W)$ defined by the equation~(\ref{C-for}).

Then, the final cumulative distribution function $F(D,W)$ is
\begin{equation}
F(D, W)=\left\{\begin{aligned}
0, \quad D\leq D_{min}+W/C \\
1-\exp\left\{-\lambda (D- D_{min}-W/C)\right\},\\ \quad D>D_{min}+W/C
\end{aligned} \right.
\label{DexpF}
\end{equation}
where 
\begin{equation}
\lambda=1/(D_{av}-D_{min}) 
\label{lambda1}
\end{equation}
is reciprocal to the difference between average network delay $D_{av}(W)=\mathbb{E}[D(W)]$ and  minimum delay $D_{min}(W)$ and $C$ is end-to-end capacity.

It is necessary to mark that control signals are told by packets of the different size that brings the additional contribution to a delay variation (network jitter $j$). And, the less available bandwidth of end-to-end connectivity, the network jitter will be stronger~\cite{drm}. Therefore, the best the controlling algorithm will form packages of the identical size. If we use the utility {\em ping \/} for delay definition in it there is a special key for resizing of a testing package ($-l$ in Windows, $-s$ in Linux).

\section{Conclusion}
\label{s7}

In the present work for the description of process of the packet delay in a global networks it has been chosen exponential distribution. In comparison with truncated normal distribution it has shown the best correlation with experimental results.

Experimental data were gathered by means of the precision RIPE measuring system to within microseconds, and also by means of the standard utility {\em ping\/}. This utility measures round-trip time to within milliseconds. During small periods about several minutes when it is possible to consider conditions of transmission on a TCP/IP network invariable, such approach gives correlations from above 0.99. At change of network conditions the elementary {\em ping\/} testing by a series from 10 packets will allow to change exponential distribution parameters instantly.

In summary we would like to thank Leonid Fridmana, the professor from University of Mexico for fruitful dialogues in which course the idea of this article has taken shape.

\end{document}